\documentclass[%
 reprint,longbibliography,preprintnumbers,
nofootinbib,
 amsmath,amssymb,
 aps,
prl,
]{revtex4-1}
\pdfoutput=1
\usepackage[utf8]{inputenc}
\usepackage{flushend}
\usepackage{dcolumn}% Align table columns on decimal point
\usepackage{bm}% bold math
\usepackage{balance}
\usepackage[normalem]{ulem}

\usepackage[colorlinks = true,
            linkcolor = blue,
            urlcolor  = blue,
            citecolor =green,
            anchorcolor = blue]{hyperref}
\usepackage{verbatim}
\usepackage{color,ulem}
\usepackage[english]{babel}

\usepackage[utf8]{inputenc}
\input Starburst.fd
\newcommand*\initfamily{\usefont{U}{Starburst}{xl}{n}}\initfamily

\newcommand{\beq}{\begin{eqnarray}}
\newcommand{\eeq}{\end{eqnarray}}
\usepackage{amsmath}
\usepackage{tikz}
\usetikzlibrary{decorations.pathmorphing}
\usetikzlibrary{shapes.misc}
\tikzset{cross/.style={cross out, draw=black, minimum size=8*(#1-\pgflinewidth), inner sep=0pt, outer sep=0pt},
cross/.default={1pt}}
\usetikzlibrary{patterns,math}
\begin{document}

\title{Can the noble metals (Au, Ag and Cu) be superconductors?}

\author{Giovanni Alberto Ummarino$^{1}$}
\author{Alessio Zaccone$^{2,3}$}

\affiliation{$^1$Politecnico di Torino, Department of Applied Science and Technology, Corso Duca degli Abruzzi 24, 10129 Torino, Italy.
}
%\address{$^2$ National Research Nuclear University MEPhI (Moscow Engineering Physics Institute), Kashira Hwy 31, Moskva 115409, Russia.}
\affiliation{$^2$Department of Physics ``A. Pontremoli'', University of Milan, via Celoria 16,
20133 Milan, Italy}
\affiliation{$^3$Institut f{\"u}r Theoretische Physik, University of G{\"o}ttingen,
Friedrich-Hund-Platz 1,
37077 G{\"o}ttingen, Germany}

 \vspace{1cm}

\begin{abstract}
It is common knowledge that noble metals are excellent conductors but do not exhibit superconductivity. On the other hand, quantum confinement in thin films has been consistently shown to induce a significant enhancement of the superconducting critical temperature in several superconductors. It is therefore an important fundamental question whether ultra-thin film confinement may induce observable superconductivity in non-superconducting metals.
We present a generalization, in the Eliashberg framework, of a BCS theory of  superconductivity in good metals under thin-film confinement. By numerically solving these new Eliashberg-type equations, we find the dependence of the superconducting critical temperature on the film thickness $L$. This parameter-free theory predicts a maximum increase in the critical temperature for a specific value of the film thickness, which is a function of the number of free carriers in the material. Exploiting this fact, we predict that ultra-thin films of gold, silver and copper of suitable thickness could be superconductors at low but experimentally accessible temperatures. We demonstrate that this is a fine-tuning problem where the thickness must assume a very precise value, close to half a nanometer.
\end{abstract}

\maketitle

%\section{Introduction}
It is well known that the three best conducting metals, Au Ag and Cu, are also among the few metallic elements that are not superconductors even when subjected to high pressures \cite{Khan1975,Buzea_2005}. In this article we  demonstrate, by exploiting the phenomenon of quantum confinement, that it is possible to make these materials superconducting as long as they are cast into ultra-thin films of a very well-defined thickness.
The superconducting critical temperatures will still be low but not so low that they cannot be measured experimentally.
The standard one-infinite-band s-wave Eliashberg theory \cite{Eliashberg,revcarbi} is a powerful tool to compute all superconductive properties of elemental superconductors \cite{revcarbi} such as as Pb, Sn, Al etc.
%Over the past few decades, however, also experimental data have appeared that seem not to be in agreement with this theory \cite{revcarbi}. These are critical temperature ($T_c$) measurements on thin lead (Pb) films as a function of thickness, which are in contrast to the behavior predicted theoretically based on the standard BCS theory.
In this regard, many studies have been devoted to rationalizing the dependence of the superconducting
critical temperature $T_{c}$ on confinement and on the thin film thickness $L$ \cite{ThompsonBlatt,Arutyunov2019,valentinis,Bianconi,lead1,lead2}.
In the past, due to the vapor-deposition technique \cite{Buckel1954}, superconducting thin films were
mostly amorphous while nowadays, thanks to the modern preparation techniques,
also crystalline thin films can be fabricated.
Early numerical studies based on BCS theory \cite{ThompsonBlatt} suggested a possible enhancement of $T_{c}$ upon decreasing the film thickness $L$, although a mechanistic explanation has remained elusive.
More recently, experiments on ordered thin films \cite{lead1,lead2,doi:10.1126/sciadv.adf5500}, besides the
above mentioned regime of $T_c$ enhancement upon reducing $L$, have
also highlighted a second regime at lower (nanometric and sub-nanometric) thicknesses $L$, where, instead, $T_{c}$
grows with increasing $L$. This behaviour results, overall, in a non-monotonic trend with a peak or maximum of $T_c$ as a function of $L$.
Travaglino and Zaccone in a recent paper \cite{zaccone}
developed the first fully analytical theory of confinement effects on
superconductivity of thin films in the framework of the simplified weak-coupling BCS formalism.
The mathematical predictions were verified
for experimental data of crystalline thin films and
were able to reproduce the trend of $T_{c}$
vs $L$, including the maximum of $T_{c}$ at $L=L_{c}=(2\pi/n)^{1/3}$, where $n$ is the concentration of free carriers. This maximum coincides with a topological transition of the Fermi surface, from the spherical-like Fermi surface of bulk metals to a non-trivial topology with homotopy group $\simeq \mathbb{Z}$. This topological transition marks the change from a situation where free electrons get crowded at the Fermi level upon decreasing the film thickness (due to the growth of hole pockets internal to the spherical Fermi surface) to a regime of strong confinement where the new topology of the Fermi surface allows for spreading out the free electron energy states at the surface. This phenomenon provides a mechanistic explanation to the maximum in $T_c$ vs thickness $L$ observed experimentally.
%In a nutshell, two symmetrical (with respect to the center
%of the Fermi sphere) spherical cavities of forbidden
%states (due to thin-film confinement) were predicted to
%grow inside the Fermi sphere upon decreasing $L$. The
%two spheres of forbidden states (hole pockets) grow further up to the
%point, at $L=L_{c}$, where the spherical Fermi surface opens up and a topological transition occurs from the
%trivial sphere $\pi_{1}(S_{2}) = 0$ to a non-trivial Fermi surface
%with homotopy group $\pi_{1}(S_{1}) = \mathbb{Z}$. In the regime
%$L > L_{c}$, as the two cavities grow with decreasing $L$, a
%redistribution of electronic states density from the interior towards
%the Fermi surface occurs, which increases the density of states (\textcolor{red}{density of states}) at the
%Fermi level. For $L < L_{c}$, instead, as $L$ decreases further,
%the Fermi surface of the system grows and the states become
%more spread out on the Fermi surface, hence the
%\textcolor{red}{density of states} at Fermi level here decreases upon further decreasing
%$L$. For Pb thin films it was found that
%$L_{c} \approx 4$ {\AA}.
%This theory is based on analytically
%describing the effect of confinement on the Fermi surface and on
%the electron density of states. 

In this paper, we formulate a generalized Eliashberg theory of strong-coupling superconductivity of noble-metal thin films, that takes into account effects of quantum confinement on the free carriers, as well as a realistic electron-phonon spectral density.
To this aim, we use a generalization of the standard s-wave one-band Eliashberg theory \cite{Eliashberg} where the new Eliashberg equations are more complex than the usual ones, because the normal density of states is not approximated by its (constant) value at the Fermi level. 
The material's physical and chemical features are taken into account in this framework via the Eliashberg spectral function $\alpha^2 F(\Omega)$. For our calculations on noble metals, we use ab-initio calculated $\alpha^2 F(\Omega)$ spectra for crystalline materials \cite{GIRI}. In the future, structural disorder effects can be taken into account by using the $\alpha^2 F(\Omega)$ spectra measured experimentally for polycrystalline or amorphous thin films \cite{Wyder}.

This  theory is shown to yield predictions for Au, Ag and Cu thin films, with no adjustable parameters. Moreover, the calculations predict that the noble metals Au, Ag and Cu, become superconductors at precise values of the film thickness $L$. These predictions have the potential to change our fundamental understanding of superconductivity in nanostructured materials, with many relevant technological applications ranging from Josephson junctions to quantum computing.
%\section{The model} Di qua in poi ho cambiato in modo sistematico prima solo in parte

The Eliashberg equations, in their simpler version (one infinite band with isotropic order parameter), are given in terms of the gap function $\Delta(i\omega_n)$ and renormalization function $Z(i\omega_n)$ \cite{revcarbi, revcarbimarsi, Allen, Parks, Marsiglio, ummarinorev,margine1}. When the Migdal's theorem holds \cite{UmmaMig}, they read as:
%
%\begin{widetext}
\begin{align}
&\Delta(i\omega_n)Z(i\omega_n)=\pi T\sum_{\omega_{n'}} \frac{\Delta(i\omega_{n'})}{\sqrt{\omega_{n'}^2+\Delta^{2}(i\omega_n)}}\times\nonumber\\
&\big[ \lambda (i\omega_{n'}-i\omega_n)-\mu^{*}(\omega_{c})\theta(\omega_{c}-|\omega_{n'}|)\big],\nonumber\\
Z(i\omega_n)&=1+\frac{\pi T}{\omega_n}\sum_{\omega_{n'}}\frac{\omega_{n'}}{\sqrt{\omega_{n'}^2+\Delta^{2}(i\omega_n)}}\lambda (i\omega_{n'}-i\omega_n)
\end{align}
%\end{widetext}
 where $\omega_{n}$ are the Matsubara energies and $n$ are integer numbers, $\mu^{*}(\omega_{c})$ is the Coulomb pseudopotential that depends, in a weak way, on a cut-off energy $\omega_{c}$ ($\omega_{c}> 3\Omega_{max}$ where $\Omega_{max}$ is the maximum phonon or Debye energy) \cite{Allen}, and $\theta(\omega_{c}-|\omega_{n'}|)$ is the Heaviside function.
$\lambda (i\omega_{n'}-i\omega_n)$ is a function related to the electron-phonon spectral function $\alpha^2F(\Omega)$ through the relation
\begin{equation} \label{lambda}
\lambda (i\omega_{n'}-i\omega_n)=2\int_0^\infty \frac{\Omega\alpha^2F(\Omega) d\Omega}{\Omega^2+(\omega_{n'}-\omega_n)^2}.
\end{equation}
The strength of the electron-phonon coupling is given by the electron-phonon coupling parameter $\lambda=2\int_0^\infty \frac{\alpha^2F(\Omega) d\Omega}{\Omega}$.
In general, it is impossible to find exact analytical solutions of Eliashberg's equations except for the case of extreme strong-coupling ($\lambda>10$) \cite{revcarbi}. Hence, we solve them numerically with an iterative method until numerical convergence is reached. This  numerical procedure is easy to perform in the formulation on the imaginary axis, but less so on the real axis.
The $T_c$ can be calculated either by solving an eigenvalue equation or, more easily, by giving a very small test value to the superconducting gap and then by checking at which temperature the solution converges. In this way, a precision in the $T_{c}$ value is obtained that is much higher than the experimental confidence interval.
The simplest thing to do to generalize the Eliashberg equations is to remove the infinite band approximation (which works very well for most metals in the bulk state) and to no longer approximate the normal density of states as a function of energy with its value at the Fermi level.
By removing these approximations, the Eliashberg equations become slightly more complex and they become four equations \cite{Allen}.
However, in the particular case where the density of states is symmetrical with respect to the Fermi level ($N(\varepsilon)=N(-\varepsilon)$), it is possible to simplify the theory in the way that the self energy terms remain just two, $Z(i\omega_n)$ and $\Delta(i\omega_n)Z(i\omega_n)$, and the equations read as \cite{carbin1,carbin2}
\begin{widetext}
\begin{align}
%\label{EliashImmFinale}
\Delta(i\omega_n)Z(i\omega_n)=&\pi T\sum_{\omega_{n'}} \frac{\Delta(i\omega_{n'})}{\sqrt{\omega_{n'}^2+\Delta^{2}(i\omega_n)}}[\frac{N(i\omega_{n'})+N(-i\omega_{n'})}{2}]\times\nonumber
%\big[\lambda(i\omega_{n'}-i\nped{\omega}{n})-\mu^{*}(\nped{\omega}{c})\theta(\nped{\omega}{c}-|\omega_{n'}|)\big]\nonumber
\\
&\big[\lambda(i\omega_{n'}-i\omega_n)-\mu^{*}(\omega_c)\theta(\omega_c-|\omega_{n'}|)\big]\frac{2}{\pi} \arctan(\frac{W}{2Z(i\omega_{n'})\sqrt{\omega_{n'}^{2}+\Delta^{2}(i\omega_{n'})}}),
\\
Z(i\omega_n)=& 1+\frac{\pi T}{\omega_n}\sum_{\omega_{n'}} \frac{\omega_{n'}}{\sqrt{\omega_{n'}^2+\Delta^{2}(i\omega_n)}}[\frac{N(i\omega_{n'})+N(-i\omega_{n'})}{2}]
\lambda (i\omega_{n'}-i\omega_n)\frac{2}{\pi}\arctan(\frac{W}{2Z(i\omega_{n'})\sqrt{\omega_{n'}^{2}+\Delta^{2}(i\omega_{n'})}})
\end{align}
\end{widetext}
where $N(\pm i\omega_{n'})=N(\pm Z(i\omega_{n'})\sqrt{(\omega_{n'})^{2}+\Delta^{2}(i\omega_{n'})})$ and the bandwidth $W$ is equal to half the Fermi energy, $E_{F}/2$.
%
%Having the \textcolor{red}{density of states} in the symmetric normal state is a great advantage for reaching solution convergence more quickly.
%In the Eliashberg theory one has to ensure that $N(\varepsilon=E_{F,bulk})=N(0)=1$, always.

When the system is confined along one of the three spatial directions, such as in thin films, the density of states features two different regimes depending on the film thickness $L$ \cite{zaccone}:
when $L>L_{c}$ and $E_{F}>\varepsilon^{*}$, the density of states has the following form
$N(\varepsilon)=N(0)C
[\theta(\varepsilon^{*}-\varepsilon)\sqrt{\frac{E_{F}}{\varepsilon^{*}}}\frac{|\varepsilon|}{E_{F}}+
\theta(\varepsilon-\varepsilon^{*})\sqrt{\frac{|\varepsilon|}{E_{F}}}]$
\noindent where $C=(1+\frac{1}{3}\frac{L_{c}^{3}}{L^{3}})^{1/3}$, $\varepsilon^{*}=\frac{2\pi^{2}\hbar^{2}}{mL^{2}}$, $L_{c}=(\frac{2\pi}{n_{0}})^{1/3}$,
$m$ is the electron mass, $L$ is the film thickness, $n_{0}$ is the density of carriers and $E_{F,bulk}$ is the Fermi energy of the bulk material. In this case, it is possible to demonstrate the following relations \cite{zaccone}:
\begin{align}
E_{F}&=C^{2}E_{F,bulk}\\
N(E_{F})&=C N(E_{F,bulk})=CN(0),
\end{align}
with $N(E_{F,bulk})=\frac{V(2m)^{3/2}}{2\pi^{2}\hbar^{3}}\sqrt{E_{F,bulk}}$.
In the regime $\epsilon < \epsilon^*$, the density of states has a new, linear dependence on the energy, in contrast with the standard square-root dependence which is retrieved for $\epsilon > \epsilon^*$ \cite{zaccone}.
%In order to better understand the new physics hidden in these equations
%we remove the factor $C$, which will be put in the renormalization of the electron-phonon coupling constant.
We can summarize the main features that change in this version of the Eliashberg equations:
i) the density of states will no longer be a constant but a function of energy;
%In order to better understand the new physics hidden in these equations
%we removed the factor $C$. This factor will be put in the renormalization of the electron-phonon coupling constant so the \textcolor{red}{density of states} for $L>L_c$ is
%\begin{equation}
%N(\varepsilon)=\left[\vartheta(\varepsilon^{*}-\varepsilon)\sqrt{\frac{E_{F}}{\varepsilon^{*}}}\frac{|\varepsilon|}{E_{F}}+
%\vartheta(\varepsilon-\varepsilon^{*})\sqrt{\frac{|\varepsilon|}{E_{F}}}\right].
%\end{equation}\\ 
ii) the electron-phonon interaction is a function of film thickness $L$, via $\lambda=C\lambda^{bulk}$; %For simplicity we rescale the electron-phonon spectral function without changing its shape. We move the prefactor of the normal \textcolor{red}{density of states}, $C$, inside the definition of electron-phonon coupling as in the Coulomb pseudopotential. This choice allows one to justify the use of Allen-Dynes equation \cite{Dynes} for $T_{c}$.\\
iii) the value of the Fermi energy is also a function of the film thickness $L$: $E_{F}=C^{2}E_{F,bulk}$. Of course, in the symmetric case discussed above, it is $W=E_{F}/2$; 
iv) the Coulomb pseudopotential $\mu^{*}$ also depends on the film thickness via
$\mu^{*}=\frac{C\mu_{bulk}}{1+\mu_{bulk}\ln(E_{F}/\omega_{c})}$
\noindent where $\mu_{bulk}=\frac{\mu^{*}_{bulk}}{1-\mu^{*}_{bulk}\ln(E_{F,bulk}/\omega_{c})}$.
Instead, when $L<L_{c}$ and $E_{F}<\varepsilon^{*}$, we have \cite{zaccone}:
\begin{equation}
N(\varepsilon)=C'N(0)\sqrt{\frac{E_{F}}{\varepsilon^{*}}}\frac{\varepsilon}{E_{F}}
\end{equation}
\noindent where
$N(\varepsilon=E_{F})=C'N(0)$, 
$E_{F}=C'^{2}E_{F,bulk}$ and 
$C'=\frac{2}{6^{1/3}}\sqrt{\frac{L}L_c}$. In this regime, the density of states is given by \cite{zaccone}:
$N(\varepsilon)=\sqrt{\frac{E_{F}}{\varepsilon^{*}}}\frac{|\varepsilon|}{E_{F}}$. The
electron-phonon coupling and the Coulomb pseudopotential become thickness-dependent through $C'$:
\begin{equation}
\lambda=C'\lambda^{bulk} ,~~~ \mu^{*}=\frac{C'\mu_{bulk}}{1+\mu_{bulk}\ln(E_{F}/\omega_{c})}.
\end{equation}
%In the Eliashberg Equations the reference energy is the Fermi energy which is the zero of the energy.\par
%In the code that numerically solves the Eliashberg equations, by recalling that the reference energy is the Fermi energy taken as the zero of the energy, the \textcolor{red}{density of states} has been rescaled in the following way
%(by also putting care that the \textcolor{red}{density of states} is continuous for $\varepsilon=\varepsilon^{*}$). When $L>L_{c}$ and $\varepsilon^{*}<E_{F}$:
%\begin{equation}
%N(\varepsilon)=[\vartheta(\varepsilon^{*}-\varepsilon)\sqrt{\frac{E_{F}}{E_{F}-\varepsilon^{*}}}(1-\frac{|\varepsilon|}{E_{F}})+
%\vartheta(\varepsilon-\varepsilon^{*})(1-\sqrt{\frac{|\varepsilon|}{E_{F}}})].
%\end{equation}
%Instead, when $L<L_{c}$ and %$\varepsilon^{*}>E_{F}$:
%\begin{equation}
%N(\varepsilon)=\sqrt{\frac{E_{F}}{\varepsilon^{*}}}(1-\frac{|\varepsilon|}{E_{F}}).
%\end{equation}
%%%%%%%%%%%%%%%%%%%%%%%%%%%%%%%%%%%%%%%%%%%%%%%%%%%
\begin{figure*}[ht]
	\centerline{\includegraphics[width=1\textwidth]{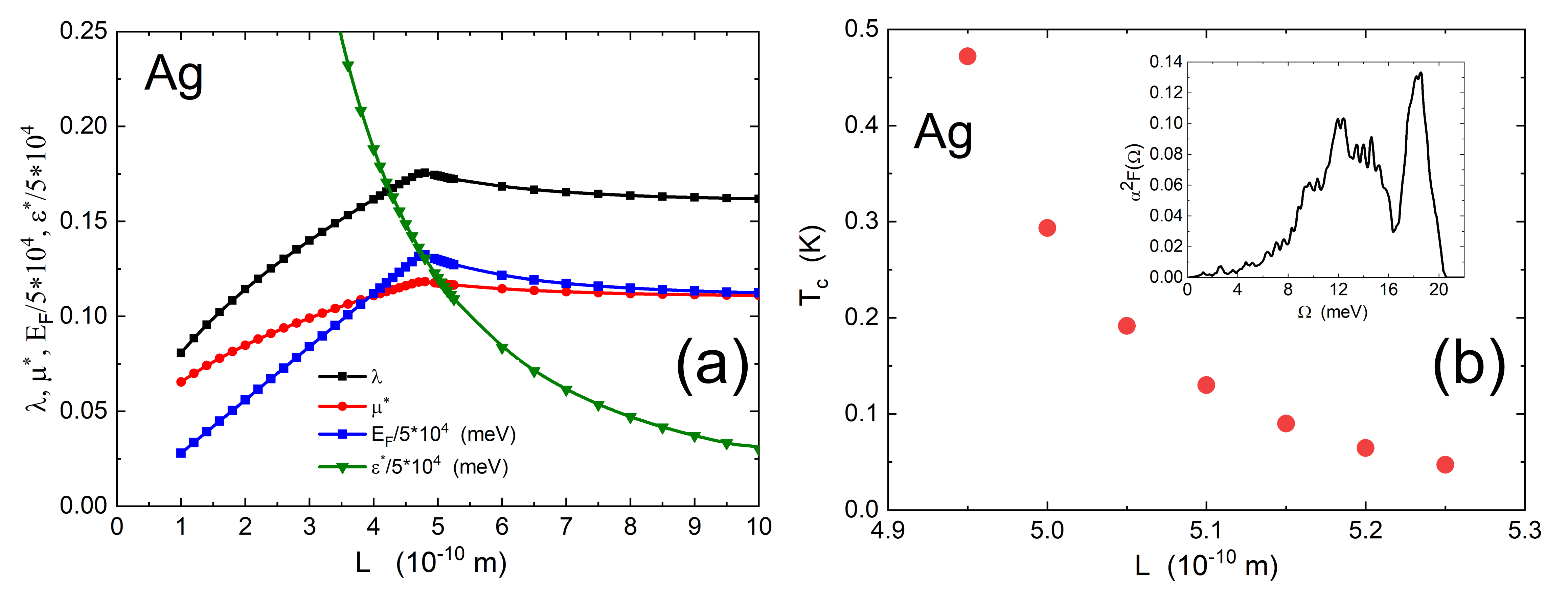}}
\caption{Panel (a) shows the physical parameters used in the theory for silver (Ag) films: $\lambda$ (dark line and squares), $\mu^{*}$ (red line and circles), $E_{F}/5\cdot 10^{4}$ (blue line and up triangles), $\varepsilon^{*}/5\cdot 10^{4}$) (green line and down triangles). All the parameters are plotted as a function of the film thickness $L$. Panel (b) shows the critical temperature $T_c$ versus film thickness $L$ for silver (Ag): full red circles represent the numerical solutions of the Eliashberg equations. In the inset, the Eliashberg electron-phonon spectral function of silver is shown, from Ref. \cite{GIRI}. }\label{diagrams1}
\end{figure*}

\begin{figure*}[t!]
	\centerline{\includegraphics[width=1\textwidth]{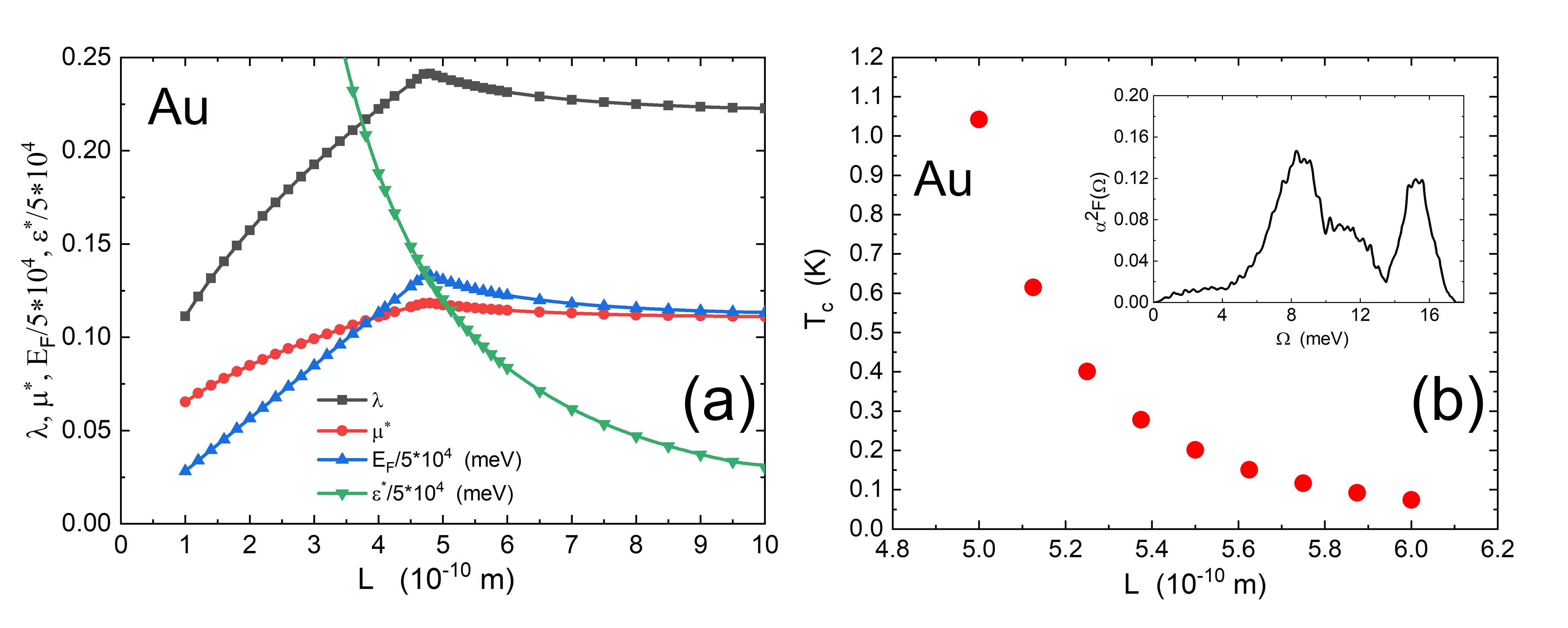}}
\caption{Panel (a) shows the physical parameters used in the theory for gold (Au) films: $\lambda$ (dark line and squares), $\mu^{*}$ (red line and circles), $E_{F}/5\cdot 10^{4}$ (blue line and up triangles), $\varepsilon^{*}/5\cdot 10^{4}$) (green line and down triangles). All parameters are plotted as a function of the film thickness $L$. Panel (b) shows the critical temperature $T_c$ versus film thickness $L$ for gold (Au): full red circles represent the numerical solutions of Eliashberg equations. In the inset, the Eliashberg electron-phonon spectral function of gold is shown, from Ref. \cite{GIRI}.}\label{diagrams2}
\end{figure*}

%\begin{figure}[htb]
%\begin{center}
%\includegraphics[keepaspectratio, width=\columnwidth]{graph2new.jpg}
%\vspace{-5mm} \caption{Critical temperature $T_c$ versus film thickness $L$ for silver (Ag): full red circles represent the numerical solutions of the Eliashberg equations. In the inset, the Eliashberg electron-phonon spectral function of silver is shown, from Ref. \cite{GIRI}. }\label{diagrams2}
%\end{center}
%\vspace{-5mm}
%\end{figure}

%\begin{figure}[htb]
%\begin{center}
%\includegraphics[keepaspectratio, width=\columnwidth]{graph3new.jpg}
%\vspace{-5mm} \caption{Physical parameters used in the theory for gold (Au) films: $\lambda$ (dark line and squares), 10$\mu^{*}$ (red line and circles), $E_{F}/10^{4}$ (blue line and up triangles), $\varepsilon^{*}/10^{4}$) (green line and down triangles). All parameters are plotted as a function of the film thickness $L$.}\label{diagrams3}
%\end{center}
%\vspace{-5mm}
%\end{figure}

%\begin{figure}[htb]
%\begin{center}
%\includegraphics[keepaspectratio, width=\columnwidth]{graph4new.jpg}
%\vspace{-5mm} \caption{$T_c$ versus film thickness $L$ for gold (Au): full red circles represent the numerical solutions of Eliashberg equations. In the inset, the Eliashberg electron-phonon spectral function of gold is shown, from Ref. \cite{GIRI}. }\label{diagrams4}
%\end{center}
%\vspace{-5mm}
%\end{figure}

\begin{figure*}[t!]	\centerline{\includegraphics[width=1\textwidth]{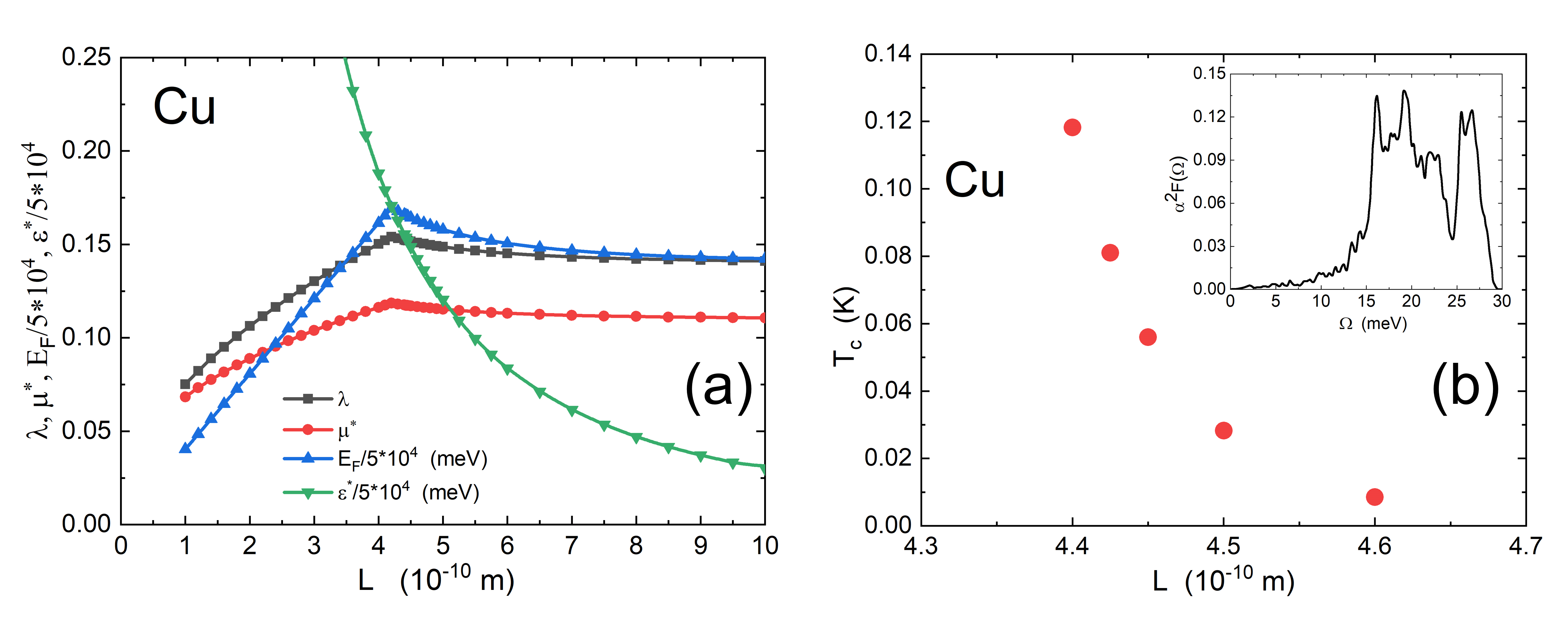}}
\caption{Panel (a) shows the values of the physical parameters used in the theory for copper (Cu) films: $\lambda$ (dark line and squares), $\mu^{*}$ (red line and circles), $E_{F}/5\cdot 10^{4}$ (blue line and up triangles), $\varepsilon^{*}/5\cdot 10^{4}$) (green line and down triangles). All the parameters are plotted as a function of the film thickness $L$. Panel (b) shows the critical temperature $T_c$ versus film thickness $L$ for copper (Cu): full red circles represent the numerical solutions of Eliashberg equations. In the inset, the Eliashberg electron-phonon spectral function of copper is shown, from Ref. \cite{GIRI}. }\label{diagrams3}
\end{figure*}

%\begin{figure}[htb]
%\begin{center}
%\includegraphics[keepaspectratio, width=\columnwidth]{graph5new.jpg}
%\vspace{-5mm} \caption{Physical parameters used in the theory for copper (cu) films: $\lambda$ (dark line and squares), 10$\mu^{*}$ (red line and circles), $E_{F}/10^{4}$ (blue line and up triangles), $\varepsilon^{*}/10^{4}$) (green line and down triangles). All the parameters are plotted as a function of the film thickness $L$.}\label{diagrams5}
%\end{center}
%\vspace{-5mm}
%\end{figure}

%\begin{figure}[htb]
%\begin{center}
%\includegraphics[keepaspectratio, width=\columnwidth]{graph6new.jpg}
%\vspace{-5mm} \caption{$T_c$ versus film thickness $L$ for copper (Cu): full red circles represent the numerical solutions of Eliashberg equations. In the inset, the Eliashberg electron-phonon spectral function of copper is shown \cite{GIRI}. }\label{diagrams6}
%\end{center}
%\vspace{-5mm}
%\end{figure}
%%%%%%%%%%%%%%%%%%%%%%%%%%%%%%%%%%%%%%%%%%%%%
%\section{Prediction on critical temperature}
We have seen that, if the normal density of states is symmetrical, the theory is simplified and we have always two Eliashberg equations to solve. Instead if the normal density of states is aymmetrical, the equations to be solved are three and the theory becomes more complex.
In general, the important thing is that, if the normal density of states is not a constant, then usually the asymmetry becomes a problem of second order.
The effect of asymmetry becomes important only in very particular situations \cite{carbidos}.
We underline that this theory is completely general because the physical and chemical properties of the specific material (including e.g. the degree of disorder, if any) are enclosed in the electron-phonon spectral function $\alpha^{2}F(\Omega)$. Of course this theory can also be easily generalized to multiband metals \cite{Umma12a,Umma12b,Umma12c}.

It is well known that all three noble metals (Au, Ag, Cu) have a very weak electron-phonon coupling ($\lambda<0.25$), which does not allow them to be bulk superconductors.
However, if we consider very thin films with a thickness very close to the critical length $L_{c}$, which is of the order of 5 {\AA} (0.5 nm), our calculations using the above theory show that the electron-phonon interaction is greatly enhanced. Therefore, the possibility exists that, in a narrow range of thickness, the noble metal film becomes superconducting.
This is the scenario revealed by our calculations in Figs. 1-3.

Let us start by considering the case of silver and examine how the fundamental parameters vary
around the critical thickness $L_c$.
In Fig. 1(a) the physical quantities of silver used in the theoretical calculations are plotted as functions of the film thickness $L$. The bulk electron-phonon spectral function of silver \cite{GIRI} with $\lambda_{bulk,0}=0.16$ is shown in the inset of Fig. 1(b). The bulk value of the Coulomb pseudopotential \cite{GIRI} is $\mu^{*}(\omega_{c})=0.11$, the cut-off energy is $\omega_{c}=75$ meV and the maximum electronic energy is $\omega_{max}=80$ meV. The values of the bulk Fermi energy and carrier density are respectively $E_{F,bulk}=5490$ meV and $n_{0}=0.0586\cdot 10^{30}$ $m^{-3}$. This produces a critical thickness $L_{c}=4.75$ {\AA}. 
As we can see from Fig. 1(a), precisely around this critical thickness value, the coupling constant $\lambda$ has a slight increase. To check whether this increase is sufficient to produce the superconducting state, we solve the modified Eliashberg equations and calculate the critical temperature $T_c$. The result is shown in Fig. 1(b).
We find that, for the film thickness $L=5.00$ {\AA} (very close to the critical value $L_{c}=4.75$ {\AA}) the material becomes a  superconductor with $T_{c}=0.294$ $K$. We notice that the thickness range that allows superconductivity to exist is quite narrow, which can be understood based on the underlying topological transition \cite{zaccone}.
%%%%%%%%%%%%%
We now turn to the case of gold.
In Fig. 2(a) the physical quantities for Au used in the theory are shown as functions of the film thickness. The bulk electron phonon spectral function of gold \cite{GIRI} with $\lambda_{bulk,0}=0.22$ is shown in the inset of Fig. 2(b). The bulk value of the Coulomb pseudopotential \cite{GIRI} is $\mu^{*}(\omega_{c})=0.11$, the cut-off energy is $\omega_{c}=55$ meV and the maximum electronic energy is $\omega_{max}=60$ meV. The values of the bulk Fermi energy and carrier density are, respectively, $E_{F,bulk}=5530$ meV and $n_{0}=0.0590\cdot 10^{30}$ $m^{-3}$. This produces a critical thickness $L_{c}=4.74$ {\AA}.
For Au, we find that, for the thickness $L=5.00$ {\AA} which is close to the critical value $L_{c}=4.74$ {\AA}, the material becomes a superconductor with $T_{c}=1.042$ $K$.
Also in this case, the thickness range that allows superconductivity to be observed is narrow.
%%%%%%%%%%%%%%%%%%%%%%%%%%
As the last case, we study copper (Cu).
In Fig. 3(a), some typical physical quantities of copper used in the theory are shown as functions of the film thickness $L$. The bulk electron-phonon spectral function of copper \cite{GIRI} with $\lambda_{bulk,0}=0.14$ is shown in the inset of Fig. 3(b). The bulk value of the Coulomb pseudopotential \cite{GIRI} is $\mu^{*}(\omega_{c})=0.11$ (the cut-off energy is $\omega_{c}=90$ meV and  the maximum electronic energy is $\omega_{max}=100$ meV). The values of the bulk Fermi energy and carrier density are, respectively, $E_{F,bulk}=7000$ meV and $n_{0}=0.0847\cdot 10^{30}$ $m^{-3}$. 
For copper we find that, if the thickness is $L=4.40$ {\AA}, i.e. close to the critical value $L_{c}=4.20$ {\AA}, the material becomes a superconductor with  $T_{c}=0.118$ $K$.

%%%%%%%%%%%%%%%%%%
%We solved the Eliashberg equations in a numerical way for the critical temperature of films of different thickness: the result is shown in Fig. 2, 4 and 6. This calculation has no free parameters.
We notice that as soon as we move away from the critical value $L_c$ of film thickness, the $T_c$ abruptly goes to very small values, which we are not able to calculate as it is too time-consuming for the code to reach convergence.

Finally, we should also point out that films that are as thin as $0.5$ nm are still effectively described by three-dimensional physics as shown plenty of times in the literature on the basis of experiments, theory and atomistic simulations, e.g. cfr. \cite{lead2,zaccone,Yu2022,Freund,Fomin}, albeit with substantial corrections due to confinement such as those implemented in our theory. 
%%%%%%%%%%%%%%%%%%%%%%%%%%%%%%%%%%%%%%%%%%%%
%\section{Conclusions}

In conclusion, we have studied a new generalization of the Eliashberg equations which include the crucial effect of quantum confinement, to compute the superconducting properties of thin films of noble metals, in a fully quantitative way and with no free parameters. Upon decreasing the film thickness, the formation of hole pockets growing inside the Fermi sea \cite{zaccone}  leads to the "crowding'' of electronic states at the Fermi level, which can significantly increase the electron-phonon coupling, and hence the $T_c$.
Surprisingly, the theoretical predictions reveal the possibility that films of Au, Ag, and Cu with a thickness close to 0.5 nm become superconducting. Particularly striking is the case of gold (Au), which can reach a superconducting critical temperature of $T_c \approx 1.1$ $K$, which is comparable to that of bulk aluminum, i.e. the most used material for Josephson junctions. These predictions open up unprecedented new avenues for both the fundamental understanding of superconductivity as well as for many technological applications, from superconducting logic to quantum computing. Also, in light of the recent experimental discovery of metallic glasses made of pure gold in Ref. \cite{Tong2024}, it will be interesting to see if an amorphous glassy structure may lead to an even greater $T_c$ by taking advantage of the more close-packed structure and of the excess of low-vibrational modes due to disorder \cite{Setty2020}. A first step in this direction would be to use Eliashberg functions of disordered thin films measured experimentally using point-contacts methods \cite{Wyder}.
Finally, the theory can be refined in future work in several respects. For example, one such angle could be to use ab-initio phonon parameters for thin films calculated with the Electron Phonon Wannier (EPW) package \cite{Margine2} (although we believe that phonon confinement should play the role of a second-order correction compared to the electronic confinement which, as demonstrated here, is responsible for a dramatic increase of the $T_c$ with increasing the confinement). Another direction could be to relax the assumption of the density of states being symmetrical with respect to the Fermi level and use instead an asymmetrical density of states, which could be relevant for certain applications \cite{Droghetti,Yakovkin}.

\subsection{Acknowledgments} 
A.Z. gratefully acknowledges funding from the European Union through Horizon Europe ERC Grant number: 101043968 ``Multimech'', from US Army Research Office through contract nr. W911NF-22-2-0256, and from the Nieders{\"a}chsische Akademie der Wissenschaften zu G{\"o}ttingen in the frame of the Gauss Professorship program. 

\bibliographystyle{apsrev4-1}

\bibliography{refs}

\begin{thebibliography}{36}%
\makeatletter
\providecommand \@ifxundefined [1]{%
 \@ifx{#1\undefined}
}%
\providecommand \@ifnum [1]{%
 \ifnum #1\expandafter \@firstoftwo
 \else \expandafter \@secondoftwo
 \fi
}%
\providecommand \@ifx [1]{%
 \ifx #1\expandafter \@firstoftwo
 \else \expandafter \@secondoftwo
 \fi
}%
\providecommand \natexlab [1]{#1}%
\providecommand \enquote  [1]{``#1''}%
\providecommand \bibnamefont  [1]{#1}%
\providecommand \bibfnamefont [1]{#1}%
\providecommand \citenamefont [1]{#1}%
\providecommand \href@noop [0]{\@secondoftwo}%
\providecommand \href [0]{\begingroup \@sanitize@url \@href}%
\providecommand \@href[1]{\@@startlink{#1}\@@href}%
\providecommand \@@href[1]{\endgroup#1\@@endlink}%
\providecommand \@sanitize@url [0]{\catcode `\\12\catcode `\$12\catcode `\&12\catcode `\#12\catcode `\^12\catcode `\_12\catcode `\%12\relax}%
\providecommand \@@startlink[1]{}%
\providecommand \@@endlink[0]{}%
\providecommand \url  [0]{\begingroup\@sanitize@url \@url }%
\providecommand \@url [1]{\endgroup\@href {#1}{\urlprefix }}%
\providecommand \urlprefix  [0]{URL }%
\providecommand \Eprint [0]{\href }%
\providecommand \doibase [0]{http://dx.doi.org/}%
\providecommand \selectlanguage [0]{\@gobble}%
\providecommand \bibinfo  [0]{\@secondoftwo}%
\providecommand \bibfield  [0]{\@secondoftwo}%
\providecommand \translation [1]{[#1]}%
\providecommand \BibitemOpen [0]{}%
\providecommand \bibitemStop [0]{}%
\providecommand \bibitemNoStop [0]{.\EOS\space}%
\providecommand \EOS [0]{\spacefactor3000\relax}%
\providecommand \BibitemShut  [1]{\csname bibitem#1\endcsname}%
\let\auto@bib@innerbib\@empty
%</preamble>
\bibitem [{\citenamefont {Khan}\ and\ \citenamefont {Raub}(1975)}]{Khan1975}%
  \BibitemOpen
  \bibfield  {author} {\bibinfo {author} {\bibfnamefont {H.~R.}\ \bibnamefont {Khan}}\ and\ \bibinfo {author} {\bibfnamefont {C.~J.}\ \bibnamefont {Raub}},\ }\href {\doibase 10.1007/BF03215082} {\bibfield  {journal} {\bibinfo  {journal} {Gold Bulletin}\ }\textbf {\bibinfo {volume} {8}},\ \bibinfo {pages} {114} (\bibinfo {year} {1975})}\BibitemShut {NoStop}%
\bibitem [{\citenamefont {Buzea}\ and\ \citenamefont {Robbie}(2004)}]{Buzea_2005}%
  \BibitemOpen
  \bibfield  {author} {\bibinfo {author} {\bibfnamefont {C.}~\bibnamefont {Buzea}}\ and\ \bibinfo {author} {\bibfnamefont {K.}~\bibnamefont {Robbie}},\ }\href {\doibase 10.1088/0953-2048/18/1/R01} {\bibfield  {journal} {\bibinfo  {journal} {Superconductor Science and Technology}\ }\textbf {\bibinfo {volume} {18}},\ \bibinfo {pages} {R1} (\bibinfo {year} {2004})}\BibitemShut {NoStop}%
\bibitem [{\citenamefont {Eliashberg}(1960)}]{Eliashberg}%
  \BibitemOpen
  \bibfield  {author} {\bibinfo {author} {\bibfnamefont {G.~M.}\ \bibnamefont {Eliashberg}},\ }\href@noop {} {\bibfield  {journal} {\bibinfo  {journal} {Sov. Phys. JETP}\ }\textbf {\bibinfo {volume} {11}},\ \bibinfo {pages} {696} (\bibinfo {year} {1960})}\BibitemShut {NoStop}%
\bibitem [{\citenamefont {Carbotte}(1990)}]{revcarbi}%
  \BibitemOpen
  \bibfield  {author} {\bibinfo {author} {\bibfnamefont {J.~P.}\ \bibnamefont {Carbotte}},\ }\href {\doibase 10.1103/RevModPhys.62.1027} {\bibfield  {journal} {\bibinfo  {journal} {Rev. Mod. Phys.}\ }\textbf {\bibinfo {volume} {62}},\ \bibinfo {pages} {1027} (\bibinfo {year} {1990})}\BibitemShut {NoStop}%
\bibitem [{\citenamefont {Blatt}\ and\ \citenamefont {Thompson}(1963)}]{ThompsonBlatt}%
  \BibitemOpen
  \bibfield  {author} {\bibinfo {author} {\bibfnamefont {J.~M.}\ \bibnamefont {Blatt}}\ and\ \bibinfo {author} {\bibfnamefont {C.~J.}\ \bibnamefont {Thompson}},\ }\href {\doibase 10.1103/PhysRevLett.10.332} {\bibfield  {journal} {\bibinfo  {journal} {Phys. Rev. Lett.}\ }\textbf {\bibinfo {volume} {10}},\ \bibinfo {pages} {332} (\bibinfo {year} {1963})}\BibitemShut {NoStop}%
\bibitem [{\citenamefont {Arutyunov}\ \emph {et~al.}(2019)\citenamefont {Arutyunov}, \citenamefont {Sedov}, \citenamefont {Golokolenov}, \citenamefont {Zav'yalov}, \citenamefont {Konstantinidis}, \citenamefont {Stavrinidis}, \citenamefont {Stavrinidis}, \citenamefont {Vasiliadis}, \citenamefont {Kekhagias}, \citenamefont {Dimitrakopulos}, \citenamefont {Komninu}, \citenamefont {Kroitoru},\ and\ \citenamefont {Shanenko}}]{Arutyunov2019}%
  \BibitemOpen
  \bibfield  {author} {\bibinfo {author} {\bibfnamefont {K.~Y.}\ \bibnamefont {Arutyunov}}, \bibinfo {author} {\bibfnamefont {E.~A.}\ \bibnamefont {Sedov}}, \bibinfo {author} {\bibfnamefont {I.~A.}\ \bibnamefont {Golokolenov}}, \bibinfo {author} {\bibfnamefont {V.~V.}\ \bibnamefont {Zav'yalov}}, \bibinfo {author} {\bibfnamefont {G.}~\bibnamefont {Konstantinidis}}, \bibinfo {author} {\bibfnamefont {A.}~\bibnamefont {Stavrinidis}}, \bibinfo {author} {\bibfnamefont {G.}~\bibnamefont {Stavrinidis}}, \bibinfo {author} {\bibfnamefont {I.}~\bibnamefont {Vasiliadis}}, \bibinfo {author} {\bibfnamefont {T.}~\bibnamefont {Kekhagias}}, \bibinfo {author} {\bibfnamefont {G.~P.}\ \bibnamefont {Dimitrakopulos}}, \bibinfo {author} {\bibfnamefont {F.}~\bibnamefont {Komninu}}, \bibinfo {author} {\bibfnamefont {M.~D.}\ \bibnamefont {Kroitoru}}, \ and\ \bibinfo {author} {\bibfnamefont {A.~A.}\ \bibnamefont {Shanenko}},\ }\href {\doibase 10.1134/S1063783419090038} {\bibfield  {journal} {\bibinfo  {journal} {Physics of the Solid
  State}\ }\textbf {\bibinfo {volume} {61}},\ \bibinfo {pages} {1559} (\bibinfo {year} {2019})}\BibitemShut {NoStop}%
\bibitem [{\citenamefont {Valentinis}\ \emph {et~al.}(2016)\citenamefont {Valentinis}, \citenamefont {van~der Marel},\ and\ \citenamefont {Berthod}}]{valentinis}%
  \BibitemOpen
  \bibfield  {author} {\bibinfo {author} {\bibfnamefont {D.}~\bibnamefont {Valentinis}}, \bibinfo {author} {\bibfnamefont {D.}~\bibnamefont {van~der Marel}}, \ and\ \bibinfo {author} {\bibfnamefont {C.}~\bibnamefont {Berthod}},\ }\href {\doibase 10.1103/PhysRevB.94.054516} {\bibfield  {journal} {\bibinfo  {journal} {Phys. Rev. B}\ }\textbf {\bibinfo {volume} {94}},\ \bibinfo {pages} {054516} (\bibinfo {year} {2016})}\BibitemShut {NoStop}%
\bibitem [{\citenamefont {Bianconi}\ and\ \citenamefont {Missori}(1994)}]{Bianconi}%
  \BibitemOpen
  \bibfield  {author} {\bibinfo {author} {\bibfnamefont {A.}~\bibnamefont {Bianconi}}\ and\ \bibinfo {author} {\bibfnamefont {M.}~\bibnamefont {Missori}},\ }\href {\doibase 10.1051/jp1:1994100} {\bibfield  {journal} {\bibinfo  {journal} {{Journal de Physique I}}\ }\textbf {\bibinfo {volume} {4}},\ \bibinfo {pages} {361} (\bibinfo {year} {1994})}\BibitemShut {NoStop}%
\bibitem [{\citenamefont {Eom}\ \emph {et~al.}(2006)\citenamefont {Eom}, \citenamefont {Qin}, \citenamefont {Chou},\ and\ \citenamefont {Shih}}]{lead1}%
  \BibitemOpen
  \bibfield  {author} {\bibinfo {author} {\bibfnamefont {D.}~\bibnamefont {Eom}}, \bibinfo {author} {\bibfnamefont {S.}~\bibnamefont {Qin}}, \bibinfo {author} {\bibfnamefont {M.-Y.}\ \bibnamefont {Chou}}, \ and\ \bibinfo {author} {\bibfnamefont {C.~K.}\ \bibnamefont {Shih}},\ }\href {\doibase 10.1103/PhysRevLett.96.027005} {\bibfield  {journal} {\bibinfo  {journal} {Phys. Rev. Lett.}\ }\textbf {\bibinfo {volume} {96}},\ \bibinfo {pages} {027005} (\bibinfo {year} {2006})}\BibitemShut {NoStop}%
\bibitem [{\citenamefont {Qin}\ \emph {et~al.}(2009)\citenamefont {Qin}, \citenamefont {Kim}, \citenamefont {Niu},\ and\ \citenamefont {Shih}}]{lead2}%
  \BibitemOpen
  \bibfield  {author} {\bibinfo {author} {\bibfnamefont {S.}~\bibnamefont {Qin}}, \bibinfo {author} {\bibfnamefont {J.}~\bibnamefont {Kim}}, \bibinfo {author} {\bibfnamefont {Q.}~\bibnamefont {Niu}}, \ and\ \bibinfo {author} {\bibfnamefont {C.-K.}\ \bibnamefont {Shih}},\ }\href@noop {} {\bibfield  {journal} {\bibinfo  {journal} {Science}\ }\textbf {\bibinfo {volume} {324}},\ \bibinfo {pages} {1314} (\bibinfo {year} {2009})}\BibitemShut {NoStop}%
\bibitem [{\citenamefont {Buckel}\ and\ \citenamefont {Hilsch}(1954)}]{Buckel1954}%
  \BibitemOpen
  \bibfield  {author} {\bibinfo {author} {\bibfnamefont {W.}~\bibnamefont {Buckel}}\ and\ \bibinfo {author} {\bibfnamefont {R.}~\bibnamefont {Hilsch}},\ }\href {\doibase 10.1007/BF01337903} {\bibfield  {journal} {\bibinfo  {journal} {Zeitschrift f{\"u}r Physik}\ }\textbf {\bibinfo {volume} {138}},\ \bibinfo {pages} {109} (\bibinfo {year} {1954})}\BibitemShut {NoStop}%
\bibitem [{\citenamefont {van Weerdenburg}\ \emph {et~al.}(2023)\citenamefont {van Weerdenburg}, \citenamefont {Kamlapure}, \citenamefont {Fyhn}, \citenamefont {Huang}, \citenamefont {van Mullekom}, \citenamefont {Steinbrecher}, \citenamefont {Krogstrup}, \citenamefont {Linder},\ and\ \citenamefont {Khajetoorians}}]{doi:10.1126/sciadv.adf5500}%
  \BibitemOpen
  \bibfield  {author} {\bibinfo {author} {\bibfnamefont {W.~M.}\ \bibnamefont {van Weerdenburg}}, \bibinfo {author} {\bibfnamefont {A.}~\bibnamefont {Kamlapure}}, \bibinfo {author} {\bibfnamefont {E.~H.}\ \bibnamefont {Fyhn}}, \bibinfo {author} {\bibfnamefont {X.}~\bibnamefont {Huang}}, \bibinfo {author} {\bibfnamefont {N.~P.}\ \bibnamefont {van Mullekom}}, \bibinfo {author} {\bibfnamefont {M.}~\bibnamefont {Steinbrecher}}, \bibinfo {author} {\bibfnamefont {P.}~\bibnamefont {Krogstrup}}, \bibinfo {author} {\bibfnamefont {J.}~\bibnamefont {Linder}}, \ and\ \bibinfo {author} {\bibfnamefont {A.~A.}\ \bibnamefont {Khajetoorians}},\ }\href {\doibase 10.1126/sciadv.adf5500} {\bibfield  {journal} {\bibinfo  {journal} {Science Advances}\ }\textbf {\bibinfo {volume} {9}},\ \bibinfo {pages} {eadf5500} (\bibinfo {year} {2023})},\ \Eprint {http://arxiv.org/abs/https://www.science.org/doi/pdf/10.1126/sciadv.adf5500} {https://www.science.org/doi/pdf/10.1126/sciadv.adf5500} \BibitemShut {NoStop}%
\bibitem [{\citenamefont {Travaglino}\ and\ \citenamefont {Zaccone}(2023)}]{zaccone}%
  \BibitemOpen
  \bibfield  {author} {\bibinfo {author} {\bibfnamefont {R.}~\bibnamefont {Travaglino}}\ and\ \bibinfo {author} {\bibfnamefont {A.}~\bibnamefont {Zaccone}},\ }\href {\doibase 10.1063/5.0132820} {\bibfield  {journal} {\bibinfo  {journal} {Journal of Applied Physics}\ }\textbf {\bibinfo {volume} {133}},\ \bibinfo {pages} {033901} (\bibinfo {year} {2023})}\BibitemShut {NoStop}%
\bibitem [{\citenamefont {Giri}\ \emph {et~al.}(2020)\citenamefont {Giri}, \citenamefont {Tokina}, \citenamefont {Prezhdo},\ and\ \citenamefont {Hopkins}}]{GIRI}%
  \BibitemOpen
  \bibfield  {author} {\bibinfo {author} {\bibfnamefont {A.}~\bibnamefont {Giri}}, \bibinfo {author} {\bibfnamefont {M.}~\bibnamefont {Tokina}}, \bibinfo {author} {\bibfnamefont {O.}~\bibnamefont {Prezhdo}}, \ and\ \bibinfo {author} {\bibfnamefont {P.}~\bibnamefont {Hopkins}},\ }\href {\doibase https://doi.org/10.1016/j.mtphys.2019.100175} {\bibfield  {journal} {\bibinfo  {journal} {Materials Today Physics}\ }\textbf {\bibinfo {volume} {12}},\ \bibinfo {pages} {100175} (\bibinfo {year} {2020})}\BibitemShut {NoStop}%
\bibitem [{\citenamefont {Jansen}\ \emph {et~al.}(1977)\citenamefont {Jansen}, \citenamefont {Mueller},\ and\ \citenamefont {Wyder}}]{Wyder}%
  \BibitemOpen
  \bibfield  {author} {\bibinfo {author} {\bibfnamefont {A.~G.~M.}\ \bibnamefont {Jansen}}, \bibinfo {author} {\bibfnamefont {F.~M.}\ \bibnamefont {Mueller}}, \ and\ \bibinfo {author} {\bibfnamefont {P.}~\bibnamefont {Wyder}},\ }\href {\doibase 10.1103/PhysRevB.16.1325} {\bibfield  {journal} {\bibinfo  {journal} {Phys. Rev. B}\ }\textbf {\bibinfo {volume} {16}},\ \bibinfo {pages} {1325} (\bibinfo {year} {1977})}\BibitemShut {NoStop}%
\bibitem [{\citenamefont {Marsiglio}\ and\ \citenamefont {Carbotte}(2008)}]{revcarbimarsi}%
  \BibitemOpen
  \bibfield  {author} {\bibinfo {author} {\bibfnamefont {F.}~\bibnamefont {Marsiglio}}\ and\ \bibinfo {author} {\bibfnamefont {J.~P.}\ \bibnamefont {Carbotte}},\ }\enquote {\bibinfo {title} {Electron-phonon superconductivity},}\ in\ \href {\doibase 10.1007/978-3-540-73253-2_3} {\emph {\bibinfo {booktitle} {Superconductivity: Conventional and Unconventional Superconductors}}},\ \bibinfo {editor} {edited by\ \bibinfo {editor} {\bibfnamefont {K.~H.}\ \bibnamefont {Bennemann}}\ and\ \bibinfo {editor} {\bibfnamefont {J.~B.}\ \bibnamefont {Ketterson}}}\ (\bibinfo  {publisher} {Springer Berlin Heidelberg},\ \bibinfo {address} {Berlin, Heidelberg},\ \bibinfo {year} {2008})\ pp.\ \bibinfo {pages} {73--162}\BibitemShut {NoStop}%
\bibitem [{\citenamefont {Allen}\ and\ \citenamefont {Mitrović}(1983)}]{Allen}%
  \BibitemOpen
  \bibfield  {author} {\bibinfo {author} {\bibfnamefont {P.~B.}\ \bibnamefont {Allen}}\ and\ \bibinfo {author} {\bibfnamefont {B.}~\bibnamefont {Mitrović}},\ }\enquote {\bibinfo {title} {Theory of superconducting tc},}\ \ (\bibinfo  {publisher} {Academic Press},\ \bibinfo {year} {1983})\ pp.\ \bibinfo {pages} {1--92}\BibitemShut {NoStop}%
\bibitem [{\citenamefont {{Parks (ed.)}}(1969)}]{Parks}%
  \BibitemOpen
  \bibfield  {author} {\bibinfo {author} {\bibfnamefont {R.~D.}\ \bibnamefont {{Parks (ed.)}}},\ }\href@noop {} {\emph {\bibinfo {title} {Superconductivity, vol. 1}}}\ (\bibinfo  {publisher} {Marcel Dekker, New York},\ \bibinfo {year} {1969})\ p.~\bibinfo {pages} {67}\BibitemShut {NoStop}%
\bibitem [{\citenamefont {Marsiglio}(1992)}]{Marsiglio}%
  \BibitemOpen
  \bibfield  {author} {\bibinfo {author} {\bibfnamefont {F.}~\bibnamefont {Marsiglio}},\ }\href {\doibase 10.1007/BF00118329} {\bibfield  {journal} {\bibinfo  {journal} {Journal of Low Temperature Physics}\ }\textbf {\bibinfo {volume} {87}},\ \bibinfo {pages} {659} (\bibinfo {year} {1992})}\BibitemShut {NoStop}%
\bibitem [{\citenamefont {Ummarino}(2013)}]{ummarinorev}%
  \BibitemOpen
  \bibfield  {author} {\bibinfo {author} {\bibfnamefont {G.}~\bibnamefont {Ummarino}},\ }\enquote {\bibinfo {title} {Eliashberg theory},}\ in\ \href@noop {} {\emph {\bibinfo {booktitle} {Emergent Phenomena in Correlated Matter}}},\ \bibinfo {editor} {edited by\ \bibinfo {editor} {\bibfnamefont {E.}~\bibnamefont {Pavarini}}, \bibinfo {editor} {\bibfnamefont {E.}~\bibnamefont {Koch}}, \ and\ \bibinfo {editor} {\bibfnamefont {U.}~\bibnamefont {Schollwoeck}}}\ (\bibinfo  {publisher} {Forschungszentrum Julich GmbH and Institute for Advanced Simulations},\ \bibinfo {address} {Julich},\ \bibinfo {year} {2013})\ pp.\ \bibinfo {pages} {13.1--13.36}\BibitemShut {NoStop}%
\bibitem [{\citenamefont {Margine}\ and\ \citenamefont {Giustino}(2013)}]{margine1}%
  \BibitemOpen
  \bibfield  {author} {\bibinfo {author} {\bibfnamefont {E.~R.}\ \bibnamefont {Margine}}\ and\ \bibinfo {author} {\bibfnamefont {F.}~\bibnamefont {Giustino}},\ }\href {\doibase 10.1103/PhysRevB.87.024505} {\bibfield  {journal} {\bibinfo  {journal} {Phys. Rev. B}\ }\textbf {\bibinfo {volume} {87}},\ \bibinfo {pages} {024505} (\bibinfo {year} {2013})}\BibitemShut {NoStop}%
\bibitem [{\citenamefont {Ummarino}\ and\ \citenamefont {Gonnelli}(1997)}]{UmmaMig}%
  \BibitemOpen
  \bibfield  {author} {\bibinfo {author} {\bibfnamefont {G.~A.}\ \bibnamefont {Ummarino}}\ and\ \bibinfo {author} {\bibfnamefont {R.~S.}\ \bibnamefont {Gonnelli}},\ }\href {\doibase 10.1103/PhysRevB.56.R14279} {\bibfield  {journal} {\bibinfo  {journal} {Phys. Rev. B}\ }\textbf {\bibinfo {volume} {56}},\ \bibinfo {pages} {R14279} (\bibinfo {year} {1997})}\BibitemShut {NoStop}%
\bibitem [{\citenamefont {Schachinger}\ and\ \citenamefont {Carbotte}(1983)}]{carbin1}%
  \BibitemOpen
  \bibfield  {author} {\bibinfo {author} {\bibfnamefont {E.}~\bibnamefont {Schachinger}}\ and\ \bibinfo {author} {\bibfnamefont {J.~P.}\ \bibnamefont {Carbotte}},\ }\href {\doibase 10.1088/0305-4608/13/12/017} {\bibfield  {journal} {\bibinfo  {journal} {Journal of Physics F: Metal Physics}\ }\textbf {\bibinfo {volume} {13}},\ \bibinfo {pages} {2615} (\bibinfo {year} {1983})}\BibitemShut {NoStop}%
\bibitem [{\citenamefont {Pickett}(1980)}]{carbin2}%
  \BibitemOpen
  \bibfield  {author} {\bibinfo {author} {\bibfnamefont {W.~E.}\ \bibnamefont {Pickett}},\ }\href {\doibase 10.1103/PhysRevB.21.3897} {\bibfield  {journal} {\bibinfo  {journal} {Phys. Rev. B}\ }\textbf {\bibinfo {volume} {21}},\ \bibinfo {pages} {3897} (\bibinfo {year} {1980})}\BibitemShut {NoStop}%
\bibitem [{\citenamefont {Mitrovi{\'c}}\ and\ \citenamefont {Carbotte}(1983)}]{carbidos}%
  \BibitemOpen
  \bibfield  {author} {\bibinfo {author} {\bibfnamefont {B.}~\bibnamefont {Mitrovi{\'c}}}\ and\ \bibinfo {author} {\bibfnamefont {J.}~\bibnamefont {Carbotte}},\ }\href@noop {} {\bibfield  {journal} {\bibinfo  {journal} {Canadian Journal of Physics}\ }\textbf {\bibinfo {volume} {61}},\ \bibinfo {pages} {784} (\bibinfo {year} {1983})}\BibitemShut {NoStop}%
\bibitem [{\citenamefont {Ghigo}\ \emph {et~al.}(2017)\citenamefont {Ghigo}, \citenamefont {Ummarino}, \citenamefont {Gozzelino},\ and\ \citenamefont {Tamegai}}]{Umma12a}%
  \BibitemOpen
  \bibfield  {author} {\bibinfo {author} {\bibfnamefont {G.}~\bibnamefont {Ghigo}}, \bibinfo {author} {\bibfnamefont {G.~A.}\ \bibnamefont {Ummarino}}, \bibinfo {author} {\bibfnamefont {L.}~\bibnamefont {Gozzelino}}, \ and\ \bibinfo {author} {\bibfnamefont {T.}~\bibnamefont {Tamegai}},\ }\href {\doibase 10.1103/PhysRevB.96.014501} {\bibfield  {journal} {\bibinfo  {journal} {Phys. Rev. B}\ }\textbf {\bibinfo {volume} {96}},\ \bibinfo {pages} {014501} (\bibinfo {year} {2017})}\BibitemShut {NoStop}%
\bibitem [{\citenamefont {Torsello}\ \emph {et~al.}(2019{\natexlab{a}})\citenamefont {Torsello}, \citenamefont {Cho}, \citenamefont {Joshi}, \citenamefont {Ghimire}, \citenamefont {Ummarino}, \citenamefont {Nusran}, \citenamefont {Tanatar}, \citenamefont {Meier}, \citenamefont {Xu}, \citenamefont {Bud'ko}, \citenamefont {Canfield}, \citenamefont {Ghigo},\ and\ \citenamefont {Prozorov}}]{Umma12b}%
  \BibitemOpen
  \bibfield  {author} {\bibinfo {author} {\bibfnamefont {D.}~\bibnamefont {Torsello}}, \bibinfo {author} {\bibfnamefont {K.}~\bibnamefont {Cho}}, \bibinfo {author} {\bibfnamefont {K.~R.}\ \bibnamefont {Joshi}}, \bibinfo {author} {\bibfnamefont {S.}~\bibnamefont {Ghimire}}, \bibinfo {author} {\bibfnamefont {G.~A.}\ \bibnamefont {Ummarino}}, \bibinfo {author} {\bibfnamefont {N.~M.}\ \bibnamefont {Nusran}}, \bibinfo {author} {\bibfnamefont {M.~A.}\ \bibnamefont {Tanatar}}, \bibinfo {author} {\bibfnamefont {W.~R.}\ \bibnamefont {Meier}}, \bibinfo {author} {\bibfnamefont {M.}~\bibnamefont {Xu}}, \bibinfo {author} {\bibfnamefont {S.~L.}\ \bibnamefont {Bud'ko}}, \bibinfo {author} {\bibfnamefont {P.~C.}\ \bibnamefont {Canfield}}, \bibinfo {author} {\bibfnamefont {G.}~\bibnamefont {Ghigo}}, \ and\ \bibinfo {author} {\bibfnamefont {R.}~\bibnamefont {Prozorov}},\ }\href {\doibase 10.1103/PhysRevB.100.094513} {\bibfield  {journal} {\bibinfo  {journal} {Phys. Rev. B}\ }\textbf {\bibinfo {volume} {100}},\ \bibinfo {pages}
  {094513} (\bibinfo {year} {2019}{\natexlab{a}})}\BibitemShut {NoStop}%
\bibitem [{\citenamefont {Torsello}\ \emph {et~al.}(2019{\natexlab{b}})\citenamefont {Torsello}, \citenamefont {Cho}, \citenamefont {Joshi}, \citenamefont {Ghimire}, \citenamefont {Ummarino}, \citenamefont {Nusran}, \citenamefont {Tanatar}, \citenamefont {Meier}, \citenamefont {Xu}, \citenamefont {Bud'ko}, \citenamefont {Canfield}, \citenamefont {Ghigo},\ and\ \citenamefont {Prozorov}}]{Umma12c}%
  \BibitemOpen
  \bibfield  {author} {\bibinfo {author} {\bibfnamefont {D.}~\bibnamefont {Torsello}}, \bibinfo {author} {\bibfnamefont {K.}~\bibnamefont {Cho}}, \bibinfo {author} {\bibfnamefont {K.~R.}\ \bibnamefont {Joshi}}, \bibinfo {author} {\bibfnamefont {S.}~\bibnamefont {Ghimire}}, \bibinfo {author} {\bibfnamefont {G.~A.}\ \bibnamefont {Ummarino}}, \bibinfo {author} {\bibfnamefont {N.~M.}\ \bibnamefont {Nusran}}, \bibinfo {author} {\bibfnamefont {M.~A.}\ \bibnamefont {Tanatar}}, \bibinfo {author} {\bibfnamefont {W.~R.}\ \bibnamefont {Meier}}, \bibinfo {author} {\bibfnamefont {M.}~\bibnamefont {Xu}}, \bibinfo {author} {\bibfnamefont {S.~L.}\ \bibnamefont {Bud'ko}}, \bibinfo {author} {\bibfnamefont {P.~C.}\ \bibnamefont {Canfield}}, \bibinfo {author} {\bibfnamefont {G.}~\bibnamefont {Ghigo}}, \ and\ \bibinfo {author} {\bibfnamefont {R.}~\bibnamefont {Prozorov}},\ }\href {\doibase 10.1103/PhysRevB.100.094513} {\bibfield  {journal} {\bibinfo  {journal} {Phys. Rev. B}\ }\textbf {\bibinfo {volume} {100}},\ \bibinfo {pages}
  {094513} (\bibinfo {year} {2019}{\natexlab{b}})}\BibitemShut {NoStop}%
\bibitem [{\citenamefont {Yu}\ \emph {et~al.}(2022)\citenamefont {Yu}, \citenamefont {Yang}, \citenamefont {Baggioli}, \citenamefont {Phillips}, \citenamefont {Zaccone}, \citenamefont {Zhang}, \citenamefont {Kajimoto}, \citenamefont {Nakamura}, \citenamefont {Yu},\ and\ \citenamefont {Hong}}]{Yu2022}%
  \BibitemOpen
  \bibfield  {author} {\bibinfo {author} {\bibfnamefont {Y.}~\bibnamefont {Yu}}, \bibinfo {author} {\bibfnamefont {C.}~\bibnamefont {Yang}}, \bibinfo {author} {\bibfnamefont {M.}~\bibnamefont {Baggioli}}, \bibinfo {author} {\bibfnamefont {A.~E.}\ \bibnamefont {Phillips}}, \bibinfo {author} {\bibfnamefont {A.}~\bibnamefont {Zaccone}}, \bibinfo {author} {\bibfnamefont {L.}~\bibnamefont {Zhang}}, \bibinfo {author} {\bibfnamefont {R.}~\bibnamefont {Kajimoto}}, \bibinfo {author} {\bibfnamefont {M.}~\bibnamefont {Nakamura}}, \bibinfo {author} {\bibfnamefont {D.}~\bibnamefont {Yu}}, \ and\ \bibinfo {author} {\bibfnamefont {L.}~\bibnamefont {Hong}},\ }\href {\doibase 10.1038/s41467-022-31349-6} {\bibfield  {journal} {\bibinfo  {journal} {Nature Communications}\ }\textbf {\bibinfo {volume} {13}},\ \bibinfo {pages} {3649} (\bibinfo {year} {2022})}\BibitemShut {NoStop}%
\bibitem [{\citenamefont {Freund}\ and\ \citenamefont {Pacchioni}(2008)}]{Freund}%
  \BibitemOpen
  \bibfield  {author} {\bibinfo {author} {\bibfnamefont {H.-J.}\ \bibnamefont {Freund}}\ and\ \bibinfo {author} {\bibfnamefont {G.}~\bibnamefont {Pacchioni}},\ }\href {\doibase 10.1039/B718768H} {\bibfield  {journal} {\bibinfo  {journal} {Chem. Soc. Rev.}\ }\textbf {\bibinfo {volume} {37}},\ \bibinfo {pages} {2224} (\bibinfo {year} {2008})}\BibitemShut {NoStop}%
\bibitem [{\citenamefont {Fomin}(2021)}]{Fomin}%
  \BibitemOpen
  \bibfield  {author} {\bibinfo {author} {\bibfnamefont {V.~M.}\ \bibnamefont {Fomin}},\ }\href {\doibase doi:10.1515/9783110575576} {\emph {\bibinfo {title} {Self-rolled Micro- and Nanoarchitectures}}}\ (\bibinfo  {publisher} {De Gruyter},\ \bibinfo {address} {Berlin, Boston},\ \bibinfo {year} {2021})\BibitemShut {NoStop}%
\bibitem [{\citenamefont {Tong}\ \emph {et~al.}(2024)\citenamefont {Tong}, \citenamefont {Zhang}, \citenamefont {Shang}, \citenamefont {Zhang}, \citenamefont {Li}, \citenamefont {Zhang}, \citenamefont {Wang}, \citenamefont {Liu}, \citenamefont {Zhao}, \citenamefont {Zhang}, \citenamefont {Ke}, \citenamefont {Zhou}, \citenamefont {Bai},\ and\ \citenamefont {Wang}}]{Tong2024}%
  \BibitemOpen
  \bibfield  {author} {\bibinfo {author} {\bibfnamefont {X.}~\bibnamefont {Tong}}, \bibinfo {author} {\bibfnamefont {Y.-E.}\ \bibnamefont {Zhang}}, \bibinfo {author} {\bibfnamefont {B.-S.}\ \bibnamefont {Shang}}, \bibinfo {author} {\bibfnamefont {H.-P.}\ \bibnamefont {Zhang}}, \bibinfo {author} {\bibfnamefont {Z.}~\bibnamefont {Li}}, \bibinfo {author} {\bibfnamefont {Y.}~\bibnamefont {Zhang}}, \bibinfo {author} {\bibfnamefont {G.}~\bibnamefont {Wang}}, \bibinfo {author} {\bibfnamefont {Y.-H.}\ \bibnamefont {Liu}}, \bibinfo {author} {\bibfnamefont {Y.}~\bibnamefont {Zhao}}, \bibinfo {author} {\bibfnamefont {B.}~\bibnamefont {Zhang}}, \bibinfo {author} {\bibfnamefont {H.-B.}\ \bibnamefont {Ke}}, \bibinfo {author} {\bibfnamefont {J.}~\bibnamefont {Zhou}}, \bibinfo {author} {\bibfnamefont {H.-Y.}\ \bibnamefont {Bai}}, \ and\ \bibinfo {author} {\bibfnamefont {W.-H.}\ \bibnamefont {Wang}},\ }\href {\doibase 10.1038/s41563-024-01967-0} {\bibfield  {journal} {\bibinfo  {journal} {Nature Materials}\ } (\bibinfo {year}
  {2024}),\ 10.1038/s41563-024-01967-0}\BibitemShut {NoStop}%
\bibitem [{\citenamefont {Baggioli}\ \emph {et~al.}(2020)\citenamefont {Baggioli}, \citenamefont {Setty},\ and\ \citenamefont {Zaccone}}]{Setty2020}%
  \BibitemOpen
  \bibfield  {author} {\bibinfo {author} {\bibfnamefont {M.}~\bibnamefont {Baggioli}}, \bibinfo {author} {\bibfnamefont {C.}~\bibnamefont {Setty}}, \ and\ \bibinfo {author} {\bibfnamefont {A.}~\bibnamefont {Zaccone}},\ }\href {\doibase 10.1103/PhysRevB.101.214502} {\bibfield  {journal} {\bibinfo  {journal} {Phys. Rev. B}\ }\textbf {\bibinfo {volume} {101}},\ \bibinfo {pages} {214502} (\bibinfo {year} {2020})}\BibitemShut {NoStop}%
\bibitem [{\citenamefont {Ponce'}\ \emph {et~al.}(2016)\citenamefont {Ponce'}, \citenamefont {Margine}, \citenamefont {Verdi},\ and\ \citenamefont {Giustino}}]{Margine2}%
  \BibitemOpen
  \bibfield  {author} {\bibinfo {author} {\bibfnamefont {S.}~\bibnamefont {Ponce'}}, \bibinfo {author} {\bibfnamefont {E.}~\bibnamefont {Margine}}, \bibinfo {author} {\bibfnamefont {C.}~\bibnamefont {Verdi}}, \ and\ \bibinfo {author} {\bibfnamefont {F.}~\bibnamefont {Giustino}},\ }\href {\doibase https://doi.org/10.1016/j.cpc.2016.07.028} {\bibfield  {journal} {\bibinfo  {journal} {Computer Physics Communications}\ }\textbf {\bibinfo {volume} {209}},\ \bibinfo {pages} {116} (\bibinfo {year} {2016})}\BibitemShut {NoStop}%
\bibitem [{\citenamefont {Droghetti}\ \emph {et~al.}(2022)\citenamefont {Droghetti}, \citenamefont {Rungger}, \citenamefont {Rubio},\ and\ \citenamefont {Tokatly}}]{Droghetti}%
  \BibitemOpen
  \bibfield  {author} {\bibinfo {author} {\bibfnamefont {A.}~\bibnamefont {Droghetti}}, \bibinfo {author} {\bibfnamefont {I.}~\bibnamefont {Rungger}}, \bibinfo {author} {\bibfnamefont {A.}~\bibnamefont {Rubio}}, \ and\ \bibinfo {author} {\bibfnamefont {I.~V.}\ \bibnamefont {Tokatly}},\ }\href {\doibase 10.1103/PhysRevB.105.024409} {\bibfield  {journal} {\bibinfo  {journal} {Phys. Rev. B}\ }\textbf {\bibinfo {volume} {105}},\ \bibinfo {pages} {024409} (\bibinfo {year} {2022})}\BibitemShut {NoStop}%
\bibitem [{\citenamefont {Yakovkin}(2018)}]{Yakovkin}%
  \BibitemOpen
  \bibfield  {author} {\bibinfo {author} {\bibfnamefont {I.~N.}\ \bibnamefont {Yakovkin}},\ }\href {\doibase https://doi.org/10.1155/2018/6919031} {\bibfield  {journal} {\bibinfo  {journal} {Advances in Condensed Matter Physics}\ }\textbf {\bibinfo {volume} {2018}},\ \bibinfo {pages} {6919031} (\bibinfo {year} {2018})},\ \Eprint {http://arxiv.org/abs/https://onlinelibrary.wiley.com/doi/pdf/10.1155/2018/6919031} {https://onlinelibrary.wiley.com/doi/pdf/10.1155/2018/6919031} \BibitemShut {NoStop}%
\end{thebibliography}%

\end{document}